\documentclass{optica-article}

\usepackage[capitalise]{cleveref}
\usepackage{hyperref}

\usepackage{comment}

\usepackage[labelformat=simple]{subcaption}

\usepackage{tikz}

\usepackage[export]{adjustbox}

\tikzset{>=latex} 
\usepackage{ifthen}
\usepackage{xcolor}
\usepackage{siunitx}

\usepackage{mathtools}

\usepackage{extarrows} 

\usepackage{float}

\journal{opticajournal} 

\articletype{Research Article}

\usepackage{lineno}

\begin{document}

\title{Real-time Terahertz Compressive Optical-Digital Neural Network Imaging}

\author{
Shao-Hsuan Wu,\authormark{1} 
Seyed Mostafa Latifi,\authormark{1,2}
Chia-Wen Lin,\authormark{3,4,5} and 
Shang-Hua Yang\authormark{1,4,5,*}}

\address{
\authormark{1}Institute of Electronics Engineering, National Tsing Hua University, Hsinchu, 30013, Taiwan \\

\authormark{2}School of Engineering,  University of Liverpool, Liverpool L69 3GH, UK \\

\authormark{3}The Institute of Communications Engineering, National Tsing Hua University, Hsinchu, Taiwan\\

\authormark{4}Department of Electrical Engineering, National Tsing Hua University, Hsinchu, 30013, Taiwan \\

\authormark{5}Terahertz Optics and Photonics Center, National Tsing Hua University, Hsinchu, 30013, Taiwan \\
}

\email{
\authormark{*}\href
{mailto:shanghua@ee.nthu.edu.tw}
{shanghua@ee.nthu.edu.tw}
} 



\begin{abstract*}
Terahertz (THz) band has recently garnered significant attention due to its exceptional capabilities in non-invasive, non-destructive sensing, and imaging applications. However, current THz imaging systems encounter substantial challenges owing to hardware limitations, which result in information loss and restricted imaging throughput during data digitization and information extraction processes. To overcome these challenges, we propose a hybrid compressive optical-digital neural network designed to facilitate both real-time THz imaging and precise object information extraction. This approach utilizes a physical encoder, an optical neural network (ONN), to transform and reduce the dimensionality of physical signals, effectively compressing them to fit the physical constraints of THz sensor arrays. After the compressed signals are captured and digitized by the THz sensor array, a jointly trained digital neural network (DNN) reconstructs the signals into their desired or original form. Our proposed THz ONN-DNN computational imaging system demonstrates enhanced imaging quality, an expanded field of view with a lens-free system, diffraction-free imaging capability, and real-time THz video capture at a rate of two frames per second.

\end{abstract*}


\section{Introduction}
Over the past decades, terahertz (THz) technology has advanced rapidly due to its non-ionizing and non-destructive nature~\cite{Walker_2002_10_safety_guidelines, Berry_2003_6_safety_issues, Clothier_2003_6_Effects}. It offers a chemically informative spectrum~\cite{jepsen2011terahertz} and can penetrate non-metallic and non-polar materials ~\cite{NAFTALY_2005_Terahertz_transmission}. These characteristics make THz technology particularly well suited for non-invasive imaging applications, facilitating its adoption in various fields, including biomedical imaging~\cite{wan2020terahertz, yang2016biomedical, zhang2020continuous}, quantum sensing and imaging~\cite{kutas2020terahertz}, chemical identification~\cite{Kawase_2003_5_Non, federici2005thz}, and circuit inspection~\cite{park2015non}. Consequently, THz imaging has become increasingly prominent in sectors like bio-informatics, industrial inspection, and security screening. The increasing demand for THz imaging has made the development of real-time, high-throughput imaging systems essential. These real-time systems not only improve the practicality of existing applications but also facilitate the capture and analysis of dynamic, time-sensitive data, opening up new possibilities across various fields.

The initial approach to THz imaging employed a two-axis mechanical stage for raster scanning, enabling signal collection from a defined region of interest~\cite{hu1995imaging}. While this method produces high-quality THz images with wavelength-scale spatial resolution~\cite{koch1998thz}, it suffers from a significant drawback: the data acquisition time scales linearly with image resolution. For example, scanning a centimeter-scale object may require several minutes to hours, with higher spatial resolutions further extending this duration.
To overcome these limitations and achieve practical imaging speeds for real-world applications, focal-plane array (FPA) sensors have been introduced. These sensors integrate multiple THz detectors into 1D or 2D array formats, allowing parallel signal acquisition and significantly accelerating the imaging process.

Common FPA sensor types for THz imaging include time-resolved sensor arrays, field-effect transistor (FET) arrays, and microbolometer arrays. Time-resolved sensor arrays typically utilize THz photoconductive antennas (PCAs) or electro-optic (EO) sensors, paired with a pulsed THz emitter and a femtosecond laser in a pump-probe measurement setup, enabling real-time imaging~\cite{li2024plasmonic, wu1996two}. However, these systems require high-power, bulky femtosecond lasers and mechanical delay stages, which considerably increase system size, cost, and complexity, thereby limiting their practical usability.
THz FET array sensors address some of these challenges by coupling incident THz waves with plasma waves between the source and drain, generating high-speed currents proportional to the THz signal strength~\cite{lisauskas2014exploration, lisauskas2009terahertz, nadar2010room}. These sensors feature compact designs and operate efficiently at room temperature. Nonetheless, scaling THz FET arrays to larger formats remains a significant fabrication challenge, and the limited space-bandwidth product continues to be a persistent issue~\cite{li2023high}.

Microbolometer FPA sensors, on the other hand, offer a higher space-bandwidth product. These sensors consist of an array of bolometric detectors that absorb incoming THz radiation and convert it into electrical signals~\cite{karasik2011nanobolometers}. Commercial microbolometer arrays can achieve resolutions as high as \(388 \times 288\) pixels~\cite{chevalier2013introducing}, enabling laser-free THz imaging at frame rates of tens of frames per second. Despite these advantages, microbolometer FPAs face a critical challenge: they are highly susceptible to diffraction-induced artifacts. These artifacts arise inherently from wave propagation and can lead to substantial information loss, particularly when the ratio of aperture size to wavelength approaches unity ($\dfrac{\text{aperture size}}{\lambda} \approx 1$). Under these conditions, diffraction effects become more pronounced, posing significant challenges for reconstructing fine interior details of objects at the millimeter scale.

In summary, while FPA-based THz imaging systems represent a significant advancement over mechanical raster scanning, their practical implementation in real-time imaging applications continues to be hindered by diffraction artifacts, scalability issues, and system complexity. Addressing these challenges remains essential for advancing the capabilities and adoption of THz imaging technology.

To balance imaging speed, image quality, and information reconstruction, several computational THz imaging approaches have been proposed~\cite{su2023physics, su2023THz_CT, hung2024terahertz}. Among these, THz compressed sensing (CS) has emerged as a powerful technique that breaks the Nyquist sampling limit~\cite{chan2008single} and significantly reduces data acquisition time compared to traditional raster scanning methods. Notably, a 6-fps THz CS video was achieved using a single-pixel detector~\cite{stantchev2020real}.
However, high-speed THz CS imaging using spatial light modulators faces inherent trade-offs between imaging speed, resolution, and field of view. Increasing the imaging area often compromises resolution due to the fixed pixel count of the modulators. This trade-off remains a critical bottleneck in advancing THz CS for practical high-resolution, large-area imaging applications.

To address these challenges, recently Lin et al. introduced a physical computational imaging approach leveraging an optical neural network (ONN) constructed from diffractive optical elements~\cite{lin2018all}. The ONN performs high-dimensional correlated functions between input and output optical wavefronts as the optical signal propagates through the diffractive layers. This setup enables certain computational tasks to be offloaded from digital processors to the optical domain, significantly reducing the computational burden on traditional computer hardware~\cite{fu2024optical}.
Despite its promise, ONN-based THz imaging systems remain heavily constrained by the sensing capabilities of THz detectors~\cite{icsil2024all, lin2018all, bai2024pyramid, li2024all}. The overall performance of the system is intrinsically tied to the sensor specifications, including sensitivity, dynamic range, and resolution, as well as the data format captured by the sensors. These dependencies highlight the need for continued advancements in both THz sensor technologies and computational imaging algorithms to fully realize the potential of ONN-enabled THz imaging systems.

Here, we propose a compressive THz computational imaging framework that synergistically integrates a hybrid optical-digital neural network (ONN-DNN). This framework harnesses the strengths of ONNs for light-speed signal processing, THz FPAs for high-speed data acquisition, and digital neural networks (DNNs) for efficient signal decompression and information retrieval. In this system, the ONN encodes high-dimensional THz signals into a compact, low-dimensional representation. This transformation reduces both hardware complexity and the sheer volume of raw data collected during acquisition. The compressed optical signals are captured by THz FPAs, enabling rapid imaging at video rates. The DNN processes and decompresses the low-dimensional signals back into high-dimensional data, reconstructing detailed object information with high fidelity. 

This hybrid imaging system offers several benefits: (1) Video-rate imaging: By leveraging light-speed processing capabilities of ONNs and the rapid data acquisition rates of THz FPAs, the system achieves imaging speeds of tens of frames per second, enabling real-time operation. (2) Reduced hardware constraints: The ONN's pre-processing capabilities minimize reliance on high-specification hardware, allowing compact and cost-efficient sensor designs without compromising imaging performance. (3) Diffraction-free imaging: The DNN effectively mitigates diffraction-induced artifacts during signal reconstruction, ensuring high-resolution image fidelity even under challenging diffraction conditions. (4) Efficient data processing: By transforming vast volumes of THz data into compact tokens, the system optimizes data transfer, storage, and processing workflows, allowing low-pixel-number THz sensor arrays to capture multi-scale information from THz signals. This hybrid framework is versatile and can seamlessly integrate with existing computational imaging methodologies and state-of-the-art 3D-printed THz device designs. Its adaptability makes it suitable for a broad range of real-world applications, including industrial inspection, security screening, material development, bioinformatics, and space exploration. In conclusion, the proposed compressive THz computational imaging framework bridges the gap between optical and digital domains, offering a scalable, efficient, and high-performance  THz imaging solution.

\section{Principle}
\label{sec:design_theory_Principle}
In this section, we introduce the working principle of optical-digital signal dimensional transformation, which serves as the foundation of our hybrid neural network THz computational imaging system. The system adopts an encoder-decoder architecture, where the NN functions as the encoder and the DNN serves as the decoder. Specifically, the ONN transforms high-dimensional THz signals into a compact, lower-dimensional latent space, encoding essential features in an optimized data format. This transformation reduces hardware complexity and data volume, enabling efficient acquisition and transmission of THz signals. Then, the DNN decompresses the THz signals penetrating an object from their low-dimensional representation back into a high-dimensional space. Through this process, the DNN reconstructs the object's spatial and spectral information with high fidelity, preserving intricate details and mitigating signal degradation. We will next illustrate the methodology for designing and identifying the encoder and decoder functions. This process involves balancing multiple objectives, including dimensionality reduction, preservation of meaningful latent representations, accurate reconstruction, and robust generalization. Additionally, it is essential to ensure that the encoder and decoder function seamlessly across various environments, both in the optical and digital domains. Following this, we will demonstrate the training steps for our proposed hybrid neural network THz computational imaging system.

\subsection{Optical-digital data encoding and decoding}
\label{subsec:Dimension compressing}
\begin{figure}[htbp]
    \centering
    \includegraphics[width=0.9\textwidth]{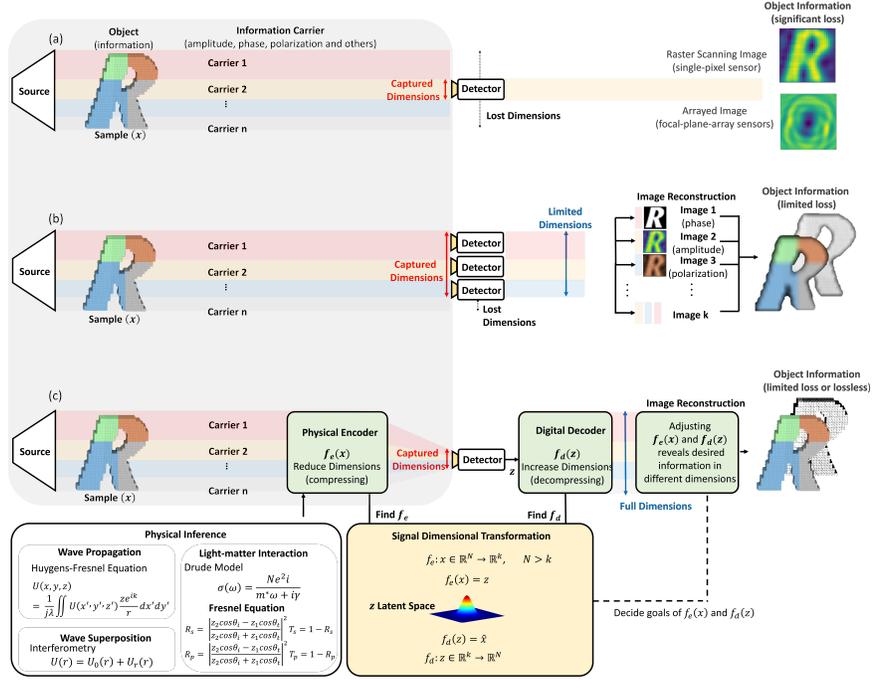}
    \caption{(a) A conventional imaging system that utilizes a sensor module to capture specific information from an optical signal carrier. (b) A computational imaging system that employs multiple sensor modules or multifunctional sensors to capture a broader range of signal dimensions. Object information is then extracted using physics-guided computational methodologies. (c) A hybrid optical-digital imaging system incorporating a physical encoder and a digital decoder. The physical encoder compresses the complete object information from full dimensions to a lower dimension, while the digital decoder decompresses this information back to full dimensions and reconstructs the object information.}
    \label{fig:concept}
\end{figure}

In our proposed ONN-DNN imaging framework, the core concept is signal dimension transformation. Here, ``signal dimension'' refers to the types of signals carried by the propagated electromagnetic wave (e.g., amplitude, phase, polarization), rather than the numerical structure of the data. In~\cref{fig:concept}(a), a conventional imaging system is depicted, where the dimensionality of the displayed image is determined by the detection capabilities of the sensor. For example, if bolometric sensors are used, only the amplitude dimension can be recorded, while the phase, polarization, and frequency dimensions could be lost. This information loss leads to a degradation in the quality and accuracy of the reconstructed object information. \cref{fig:concept}(b) illustrates a computational imaging system that can capture a broader range of signal dimensions through multiple sensor modules or multifunctional sensors. For instance, when using THz PCA as the sensor, it records the amplitude, phase, and polarization dimensions across the time, space, and frequency domains. The object information is then reconstructed using physics-guided computational methodologies. However, capturing additional data dimensions results in a rapid increase in data size and complexity, significantly complicating the problem of reconstructing object information. As a result, improvements in image quality are often achieved at the cost of heavy computational load and extended processing times. In~\cref{fig:concept}(c), we represent the architecture of our ONN-DNN imaging system. An optical encoder is designed to preprocess the wave characteristics---phase, amplitude, and polarization---across the time, space, and frequency domains, enabling the construction of a transformation function,
\[
f_e: x \in \mathbb{R}^N \to \mathbb{R}^K.
\] 
where \(f_e\) encodes an $N$-D signal into a $K$-D representation where $K<N$, to align the signal with the conversion limitations of the sensor while retaining essential information.

As a result, the information obtained by the sensor converts signals into digital forms, and this data-capturing process is referred to as compressed imaging. Mathematically, this is represented in a latent space (\(z\)).
A decoder is then used to decode captured digital data (\(z\)) to reconstruct data that represent and display the original $N$-D information, which can be expressed as:
\[
f_d: z \in \mathbb{R}^K \to \mathbb{R}^N.
\] 

Through this optical-digital signal transformation, our framework overcomes the limitations of conventional sensors, minimizing information loss during the conversion into digital data and enabling higher-dimensional data representation.
In the following subsections, we will provide further details on determining suitable encoder and decoder functions, along with their mathematical models, within the context of the proposed imaging framework.

\subsection{Hybrid optical-digital neural network: Design}
\label{subsec:VAE}

\subsubsection{Optical neural network}
\label{subsub:onn}
An ONN is used to process the optical beam characteristics spatially, temporally, or spectrally to perform physical computations. As an optical processor, ONN is designed to control the amplitude, phase, or polarization of the optical beam through wave propagation, wave superposition, and light-matter interactions at light speed. This primarily aims to achieve the encoded wavefront profiles along the propagation path.

To simulate the interference and propagation of light, the angular spectrum method (ASM) is employed. ASM models the optical field as a superposition of infinitely many point sources, each radiating outward according to the principles of optical field propagation. This approach allows for the calculation of the optical field distribution after propagation, expressed as

\begin{equation}
\label{eqn:propagation}
\begin{aligned}
E(x,y,z) 
= 
\iint_{-\infty}^{\infty} 
\hat{E}(k_{x}, k_{y}; 0)e^{-i[k_{x}x+k_{y}y \pm k_{z}z]} k_{x}  k_{y},
\\
\end{aligned}
\end{equation}
where $E(x,y,z)$ denotes the optical field at a specific point $(x,y,z)$ in space, $z$ is the propagation axis,  $k_{z}$ denotes the longitudinal wavenumber, and the exponential term $e^{\pm ik_{z}z}$ accounts for the phase evolution of the wave during propagation along the $z$-axis.
The term $\hat{E}(k_{x}, k_{y}; 0)$  corresponds to the 2D Fourier transform of the complex optical field $E(x,y,z)$ at the initial plane $z=0$, and $\hat{E}(k_{x}, k_{y}; z)$ can be expressed as
\begin{equation}
\label{eqn:propagation_FT}
\begin{aligned}
\hat{E}(k_{x}, k_{y}; z) 
& = \mathcal{F}\{
E(x,y,z)\}
\\
& = \frac{1}{4\pi^{2}}\iint_{-\infty}^{\infty} 
E(x,y,z)e^{i[k_{x}x+k_{y}y \pm k_{z}z]}dx .
\end{aligned}
\end{equation}

This Fourier transform describes the optical field in the spatial frequency domain, where $k_{x}$ and $k_{y}$ represent the spatial frequencies along the $x$- and $y$-axes, respectively, and $k_{z}$ captures the propagation of waves along the $z$-axis.
Where $k_{z}$ is the longitudinal wavenumber, and the exponential term $e^{\pm ik_{z}z}$ accounts for the phase evolution of the wave during propagation along the $z$-axis.
The inverse Fourier transform, used to reconstruct the optical field from its angular spectrum, is expressed as
\begin{equation}
\label{eqn:propagation_iFT1}
\begin{aligned}
E(x, y, z) 
& = \iint_{-\infty}^{\infty} 
\hat{E}(k_{x},k_{y};z)e^{-i[k_{x}x+k_{y}y]}d k_{x} d k_{y}.
\end{aligned}
\end{equation}

Thus, the angular spectrum $\hat{E}$ after propagating a distance $z$ can be related to its initial form at $z=0$ by
\begin{equation}
\label{eqn:propagation_iFT2}
\begin{aligned}
\hat{E}(k_{x}, k_{y}; z)
= \hat{E}(k_{x}, k_{y}; 0)e^{\pm k_{z}z},
\end{aligned}
\end{equation}
where the relationship between $k_{x}$, $k_{y}$, and $k_{z}$ is governed by the Helmholtz equation, $(\nabla^2 + k^2) E(x, y, z) = 0$, where $k=(\omega/c)n$ is the total wavenumber with $\omega$ being the angular frequency, $c$ the speed of light, and $n$ the refractive index of the medium.
The longitudinal wavenumber $k_{z}$ can be expressed as
\begin{equation}
\label{eqn:propagation_K}
\begin{aligned}
k_z \equiv \sqrt{k^2 - k_x^2 - k_y^2} \quad \text{with} \quad \mathrm{Im}\{k_z\} \geq 0.
\end{aligned}
\end{equation}
If $k_z$ becomes purely imaginary, the wave is evanescent, decaying exponentially along the $z$-axis while propagating in the transverse directions.

Utilizing~\cref{eqn:propagation} to~\cref{eqn:propagation_K}, the optical field after passing through an ONN can be calculated.  
The ONN utilized in our work comprises several diffractive layers, each being characterized by a unique transmittance function \(P_l(x, y)\), which governs the modulation of the optical field at the $l$-th layer. This transmittance function can be expressed as

\begin{equation}
\label{eqn:diffractive-layers-transmittance}
P_l(x, y) = \exp\left[i k \left(n(\lambda) + i \kappa(\lambda) - n_r\right) T_l(x, y)\right],
\end{equation}
where $n(\lambda)$ and $\kappa(\lambda)$ are the real and imaginary parts of the refractive index of the ONN medium, $n_r$ is the refractive index of the surrounding medium, and $T_l(x, y)$ is the thickness profile of the layer. These thicknesses are optimized using deep learning to achieve the desired optical functionality.

The total optical field after propagating through $m$ diffractive layers is given by

\begin{equation}
\label{eqn:diffractive-layers-transmittance_M}
U_m(x, y) 
 = 
\mathcal{F}^{-1}\left\{\mathcal{F}\{U_0(x, y)\} \prod_{l=1}^m \left[P_l(x, y) e^{i k_z z_\mathrm{gap}}\right]\right\},
\end{equation}
where $U_0(x, y)$ is the initial optical field, $P_l(x, y)$ represents the transmittance of the $l$-th layer, and $z_\mathrm{gap}$ is the distance between two layers.
The field at the final plane, $U_\mathrm{final}(x, y)$, is detected and represents the output of the ONN.

In short, ONNs utilize carefully designed diffractive layers and the principles of light propagation to execute complex optical computations at unparalleled speed---literally at the speed of light.

\subsubsection{Join optimization of optical neural network and digital neural network}
\label{subsub:Co-design}
In our ONN-DNN imaging system, the ONN acts as an optical encoder, compressing the high-dimensional THz signal emitted by the THz source into a latent representation in the form of a low-dimensional THz signal. This encoded signal then traverses an opaque object and is subsequently captured by the THz detector. The detected signal is digitized and processed by the DNN, which reconstructs it into a full-dimensional image.
As the encoded THz signal propagates through the object, it undergoes degradation due to absorption, diffraction, and noise introduced during the imaging process. These factors introduce randomness and uncertainty into the received signal, effectively modeling the path from the ONN to the DNN as a degradation channel with additive Gaussian noise.
To accurately recover the original object information, the DNN must not only decompress the encoded signal but also compensate for these degradations and denoise the received data. This task constitutes an ill-posed signal restoration problem, requiring robust optimization strategies.
The overall ONN-DNN framework---where the ONN encodes the THz signal into a latent space, introduces stochastic degradation, and the DNN decodes the signal back into a meaningful image---bears structural similarities to Variational Autoencoders (VAEs). Consequently, the joint training of the ONN and DNN can be formulated as an optimization problem within the VAE framework.

As mentioned above, to determine the optimal optical ONN-based encoding ($f_e(\cdot)$) and digital DNN-based decoding ($f_d(\cdot)$) functions for our ONN-DNN imaging system, we employed the training methodology of VAEs. VAEs have emerged as one of the most prominent approaches in unsupervised learning for modeling complex, high-dimensional data distributions. As part of the family of probabilistic graphical models, VAEs employ variational Bayesian methods to learn latent variable representations of data distributions~\cite{pinheiro2021variational}. 
VAEs are highly suited for real-time THz imaging due to their ability to efficiently process multi-dimensional data.
VAEs compress complex spatial, spectral, and temporal information into a compact latent space, enabling high-quality image reconstruction from sparse measurements, reducing the need for high-dimensional sensors.
Their probabilistic framework enhances robustness against noise and uncertainty, ensuring accurate object analysis while simplifying hardware requirements by compensating for missing data computationally.
Compared to other models, explicit probabilistic of VAEs modeling makes them more reliable and interpretable, addressing the key challenges of THz imaging with efficiency and precision.

The fundamental goal of a VAE is to maximize the marginal likelihood of the observed data \( x \), denoted as \( p_{\theta}(x) \). This marginal likelihood is expressed as:

\begin{equation}
\label{eqn:vae_marginal}
p_{\theta}(x) = \int p_{\theta}(x|z) p(z) \, dz,
\end{equation}
where \( z \) represents the latent variable generated by the encoder, and \( p(z) \) denotes the prior distribution over the latent variable. The term \( p_{\theta}(x|z) \) represents the conditional distribution of the observed data \( x \) given the latent variable \( z \), parameterized by \( \theta \), which corresponds to the parameters of decoder.
Significantly, the dimensionality of \( z \)  is designed to be smaller than that of \( x \), enabling a compact representation that aids in efficient processing and storage.
Since the marginal likelihood in Eq. \eqref{eqn:vae_marginal} is intractable to compute directly due to the integral over the latent variable \( z \), VAEs utilize an approximate posterior distribution, denoted \( q_{\phi}(z|x) \), to approximate the true posterior \( p_{\theta}(z|x) \). This variational approximation enables efficient training of VAEs using stochastic optimization techniques.

Training a VAE involves minimizing a total loss function that combines two terms: the reconstruction error and the Kullback-Leibler (KL) divergence between the approximate posterior \( q_{\phi}(z|x) \) and the prior \( p(z) \). The total loss function is defined as:
\begin{equation}
\label{eqn:vae_loss}
\mathcal{L}(\theta, \phi; x) = - \mathbb{E}_{q_{\phi}(z|x)}[\log p_{\theta}(x|z)] + \mathrm{KL}(q_{\phi}(z|x) \| p(z)),
\end{equation}
where the first term, \( - \mathbb{E}_{q_{\phi}(z|x)}[\log p_{\theta}(x|z)] \), represents the reconstruction error, which measures how well the decoder can reconstruct the input data \( x \) from the latent variable \( z \). The second term, \( \mathrm{KL}(q_{\phi}(z|x) \| p(z)) \), enforces regularization by ensuring that the approximate posterior remains close to the prior distribution, preventing overfitting and ensuring that the latent space adheres to the structure assumed by the prior.

During each back propagation step, the parameters \( \theta \) (decoder parameters) and \( \phi \) (encoder parameters) are updated to minimize this total loss, and this optimization can be expressed as:
\begin{equation}
\label{eqn:arg_vae}
\begin{aligned}
& \theta^*, \phi^* = \arg \min_{\theta, \phi} \mathcal{L}(\theta, \phi; x),
\\
& \theta \leftarrow \theta - \eta \frac{\partial \mathcal{L}(\theta, \phi; x)}{\partial \theta}, \quad \phi \leftarrow \phi - \eta \frac{\partial \mathcal{L}(\theta, \phi; x)}{\partial \phi}
\end{aligned}
\end{equation}
where \( \theta^* \) and \( \phi^* \) are the optimal weights of the encoder and decoder, achieving the lowest loss.
The partial derivatives of \( \theta^* \) and \( \phi^* \) are mutually dependent, and thus the model trains both parameters jointly, highlighting the joint training process of the VAE.
Through this optimization process, the model learns a latent space representation of the data that balances accurate reconstruction with a smooth, regularized latent space.

\subsection{Implementation of the proposed hybrid ONN-DNN network}
\label{subsec:training process}
In~\cref{subsub:Co-design}, we presented a mathematical training methodology designed to jointly optimize the ONN and DNN as an encoder and a decoder, respectively. This approach ensures that both networks can effectively encode and decode data while preserving the original information's statistical distribution.
To enable this joint training, we developed an optical-digital simulation environment grounded in the theoretical framework provided by the Angular Spectrum Method (ASM). This framework accurately models beam propagation and ONN operations, capturing interactions among diffractive layers, phase modulation, and absorption effects.
The simulation environment was implemented in Python (v3.8.18) with a spatial resolution set by pixel dimensions corresponding to half the operational wavelength. This subwavelength precision enables accurate modeling of intricate THz beam interactions, ensuring reliable network training under realistic conditions.

\begin{figure}[htbp]
    \centering
    \includegraphics[width=1\textwidth]{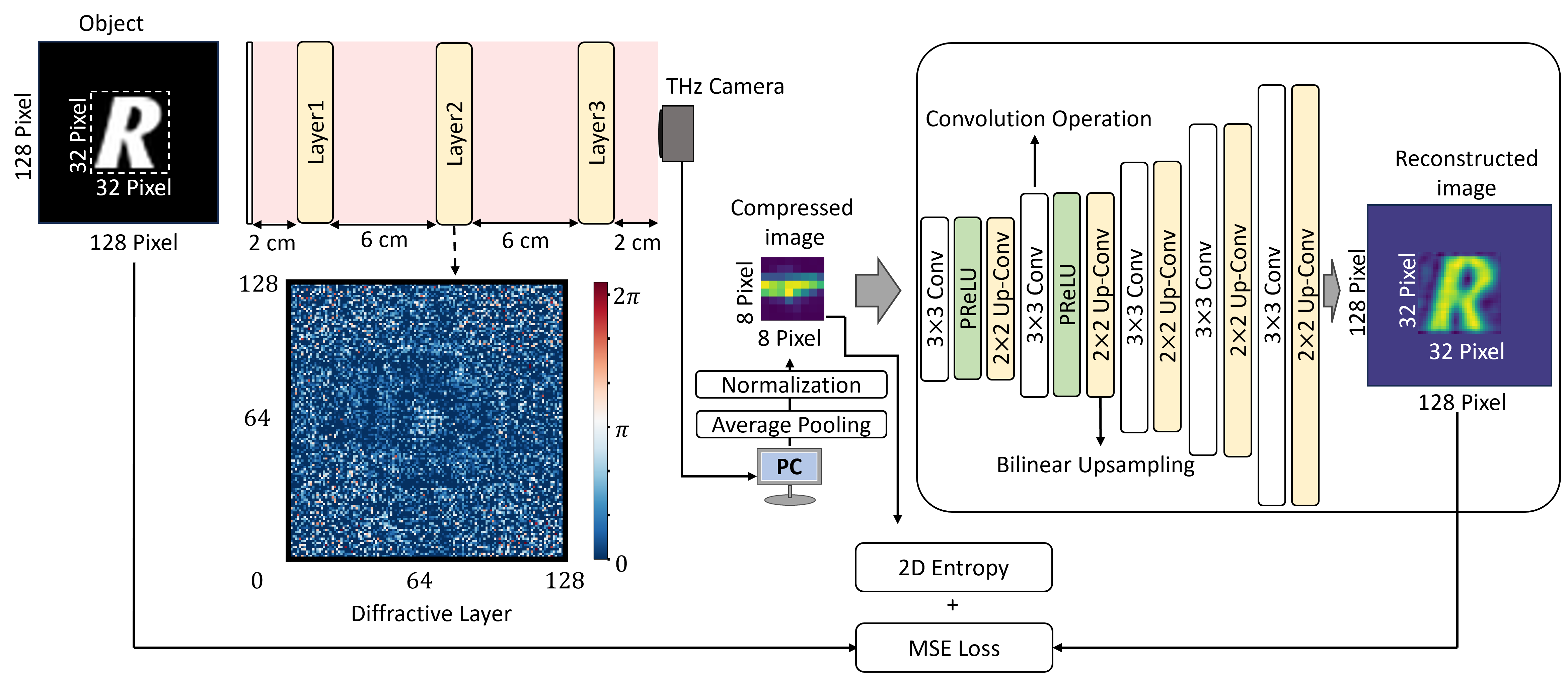}
    \caption{Simulation framework of the proposed ONN-DNN imaging system. In the optical domain, the angular spectrum method is employed to model the optical beam profile along the wave propagation axis. The setup parameters include a distance of 2 cm (13.334 $\lambda$) between the positions of the tested objects and the first diffractive layer, a spacing of 6 cm (40 $\lambda$) between each diffractive layer, and a 2 cm gap between the third diffractive layer and the THz sensor array. The THz sensor array captures the compressed optical signal and converts it to digital data for processing in the digital domain. A DNN is then utilized to decode the captured data and reconstruct the object information. The optimization of the optical and digital neural network weights is achieved through the 2D entropy of the compressed imaging and the mean squared error (MSE) loss between the input (object) and output (reconstructed image), which helps calculate the error value. ($\lambda$ = 1.5 mm)}
    \label{fig:your_label}
\end{figure}

\cref{fig:your_label} provides an overview of the simulation framework for the proposed hybrid ONN-DNN THz imaging system. The system's input is randomly selected from either the training or testing dataset, consisting of grayscale images sized \(32 \times 32\) pixels. To mitigate edge effects and enhance the stability of the propagation model, the input data is zero-padded with 48 pixels on all sides, resulting in a total input size of \(128 \times 128\) pixels. 
The ONN consists of multiple diffractive layers, each represented by trainable weight matrices of size \(128 \times 128\) pixels.
These layers operate with a spatial resolution of half the operational wavelength per pixel, enabling precise wavefront modulation and fine-grained control over beam propagation.
The primary function of the ONN is to encode and compress the input signal into the central region of the resulting beam. This spatial compression ensures alignment with the effective detection area of real-world THz focal plane array (FPA) sensors. The THz sensor operates within a physical detection area of \(12 \times 10~\mathrm{mm}\), corresponding to an effective pixel array of approximately \(16 \times 13\) pixels.
By leveraging the spatial compression achieved by the diffractive layers, the system efficiently captures essential signal information while minimizing redundancy. This approach reduces the computational overhead of the hybrid architecture and optimizes the trade-off between data processing demands and system performance.
This simulation-based training framework serves as a robust platform for advancing the capabilities of hybrid ONN-DNN models.

As the THz sensor array converts THz waves into electric signals, these signals can be digitally processed. Once converted, average pooling is applied to downsample the image to an \(8 \times 8\) pixel resolution.
Following downsampling, the image is normalized to a range of 0 to 1.
We refer to this downsampled, normalized image as the compressed image.
The compressed image is then input into a DNN to decode and reconstruct the original image. This DNN comprises convolutional layers, bilinear upsampling layers, and activation functions to achieve nonlinear transformations.
In the loss function, mean square error (MSE) is used to measure the discrepancy between the central $32 \times 32$ pixels of the original image and those of the reconstructed image.
Additionally, to avoid extreme sparsity that would impose excessively high alignment accuracy requirements, 2D entropy is employed to concentrate the signal within the compressed image.
The total loss function is therefore defined as the sum of MSE and 2D entropy.
After calculating the error value, the Adam optimizer, with a learning rate of 0.0001, is used to compute the gradients of the weights and update them accordingly.
The model is trained over 4000 epochs with a batch size of 8 and a weight decay of 0.1.

\section{Experimental Results}
\label{sec:design_theory_Experiment}

This study comprises two critical stages: validation in a simulated THz environment and implementation in a real-world THz system.
In the first stage, the ONN and DNN undergo joint training within a simulated, idealized environment. This phase focuses on learning an effective latent representation and achieving accurate reconstruction under controlled conditions. During this stage, key parameters within both the ONN and DNN architectures are carefully optimized to ensure maximum performance.

In the second stage, the insights and results gained from the first-stage simulation are adapted to a real-world THz system implementation. Unlike the controlled virtual environment, real-world experiments introduce additional complexities, such as non-ideal characteristics of ONN devices, inconsistencies in the performance of the THz sensor array, and alignment imperfections of THz waves. These factors can significantly affect system performance and, in some cases, introduce adverse effects.
To address these challenges, the study systematically investigates optimal system configurations, applies strategies to mitigate uncertainties within the THz sensor array, and demonstrates the feasibility and effectiveness of the proposed hybrid ONN-DNN computational imaging system under practical conditions.

\subsection{First-stage simulation results for our hybrid ONN-DNN imaging system}
\label{subsec:simulation experiments}
In~\cref{subsec:training process}, we introduced the framework of the proposed hybrid ONN-DNN imaging system, and in~\cref{subsec:VAE}, we detailed the optimization method for the model. For the training process, we utilized a dataset consisting of 6,656 uppercase alphabet letter images (A--Z), with an equal number of samples per category, sourced from Kaggle~\cite{kaggle_datasets}. Each image was preprocessed and standardized to 
$32 \times 32$ pixels.
To assess the generalization ability of our model, we employed a separate dataset containing lowercase letters (26 categories, 8 samples per category). This dataset provided insights into the model's capacity to handle previously unseen data during training.

While~\cref{subsec:training process} described key system parameters, such as the learning rate and optimization configurations, these parameters are generally robust across different scenarios and have minimal impact on training performance in specialized contexts. Therefore, we adopted optimal values established in prior studies.

However, several parameters significantly influence the training outcomes and overall system efficiency. To ensure optimal performance, we systematically trained the model under various configurations, focusing on identifying the most effective parameter combinations. The primary factors investigated in these experiments include:
(1) The number of diffractive layers: varying the number of layers to determine their impact on encoding and compression efficiency in the hybrid system. 
(2) The number of convolutional layers in DNN: assessing how the depth of the DNN affects its capacity to generalize and complement the optical components.
(3) The channel number of convolutional layers in DNN: exploring the impact of channel size on feature extraction and representation capacity.
(4) Activation functions in the DNN: examining the effect of incorporating activation functions versus using linear transformations only.

Through these parameter experiments, we aimed to identify the optimal architectural configurations that balance model performance and computational efficiency. The insights derived from these simulations highlight key trade-offs and synergies among the architectural components of the hybrid ONN-DNN model, offering a foundation for further optimization and real-world implementation.

\begin{figure}[htbp]
    \centering
    \includegraphics[width=0.9\textwidth]{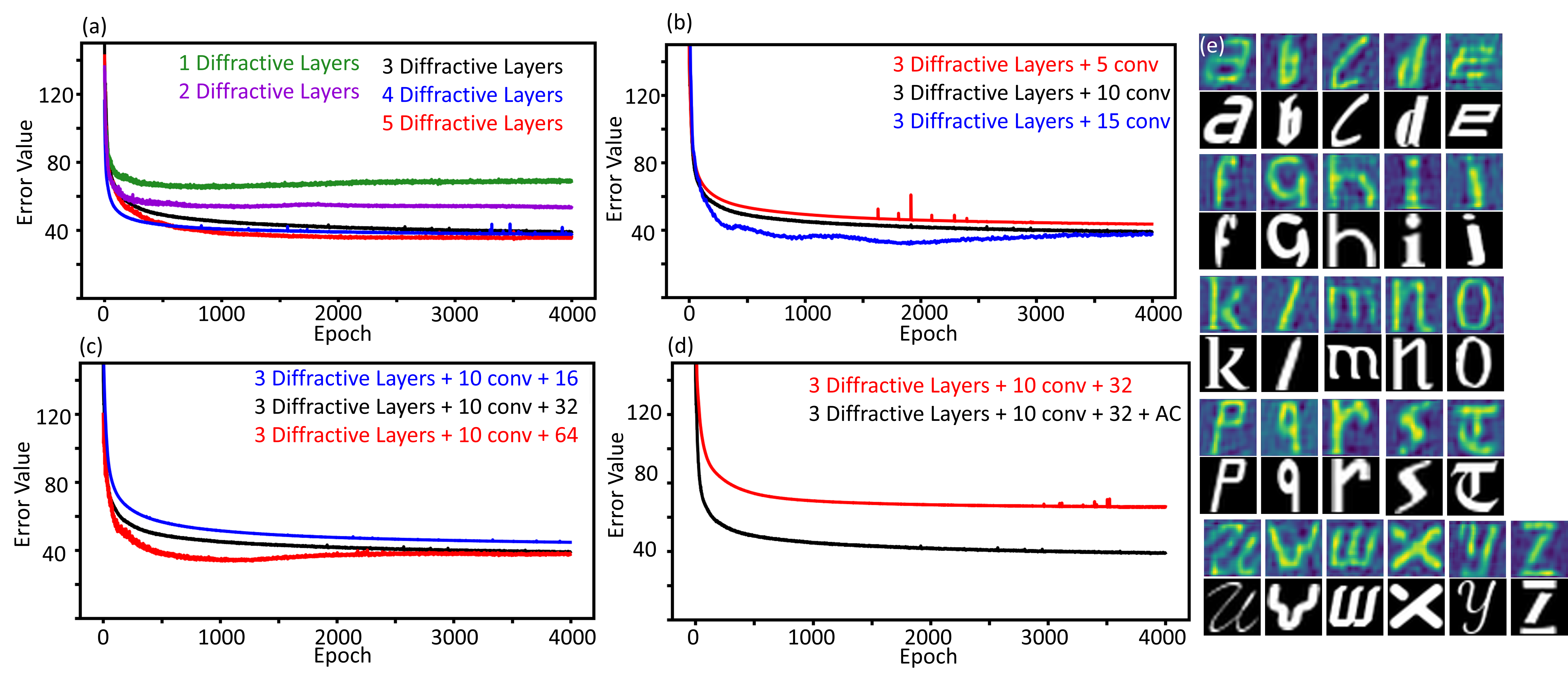}
    \caption{Comparison of error values based on: (a) the number of diffractive layers (1 to 5) in the ONN; (b) the number of convolutional layers (5, 10, and 15) with 3 diffractive layers; (c) the number of channels per convolutional layer (16, 32, and 64) with 3 diffractive layers and 10 convolutional layers; (d) the impact of including an activation function (AC) versus no activation function with 3 diffractive layers, 10 convolutional layers, and 32 channels per layer. (e) Simulation results for the model featuring 3 diffractive layers, 10 convolutional layers, 32 channels per layer, and an activation function.}
    \label{fig:onn_dnn_parameters}
\end{figure}

In~\cref{fig:onn_dnn_parameters}, the results of these experiments are summarized. 
~\cref{fig:onn_dnn_parameters}(a) illustrates the test performance with various numbers of diffractive layers in the ONN, ranging from 1 to 5 layers.
Beyond 5 layers, increased absorption would degrade the system signal-to-noise ratio (SNR), leading to a decrease in the quality of compressed images captured by the THz sensor array. Therefore, we do not extend beyond this point. The results indicate that the performance across 3--5 configurations is relatively close, with the configurations using 4 and 5 diffractive layers achieving slightly better performances.
However, the improvement is not substantial.
Additionally, considering the potential challenges posed by increased absorption and the alignment complexity in real-world experiments with additional diffractive layers, we opted for a configuration with 3 diffractive layers to balance performance and practical implementation feasibility.

Next, we evaluated the impact of varying the number of convolutional layers in the DNN. As shown in~\cref{fig:onn_dnn_parameters}(b), increasing the number of convolutional layers generally enhances performance due to the increased capacity of deeper networks to model complex functions and intricate relationships. However, while configurations with 15 convolutional layers exhibit slightly better performance compared to those with 10 layers, they show signs of overfitting after approximately 2000 training epochs. Furthermore, deeper networks significantly increase computational overhead, with each additional layer adding around 0.03 seconds per operation. Given the system's intended real-time imaging application, balancing computational efficiency with performance is essential. Based on these findings, we selected 10 convolutional layers for the DNN.

We also examined the effect of varying the number of channels in each convolutional layer. As depicted in~\cref{fig:onn_dnn_parameters}(c), increasing the number of channels enhances performance by expanding the network's capacity to extract and represent diverse features from input data. However, configurations with 64 channels exhibited a tendency toward overfitting and failed to deliver substantial improvements over those with 32 channels. Therefore, we opted for 32 channels per convolutional layer to strike an optimal balance between representation capacity and generalization performance.

To further evaluate the impact of nonlinearity in the model,~\cref{fig:onn_dnn_parameters}(d) compares the performance of a nonlinear model (using activation functions, AC) and a linear model (without AC). The results clearly indicate that the nonlinear model significantly outperforms its linear counterpart, highlighting the importance of activation functions in enhancing the network's representational power and learning efficiency.

The final configuration of the hybrid neural network comprises 3 diffractive layers in the ONN, 10 convolutional layers in the DNN, 32 channels per convolutional layer, and activation functions in the DNN. This optimized setup strikes an effective balance between robust performance and computational efficiency, making it well-suited for real-time imaging applications.
\cref{fig:onn_dnn_parameters}(e) presents the reconstructed test images generated by the optimized hybrid optical-digital neural network using the aforementioned configuration. The figure displays 26 test images of lowercase alphabet letters alongside their corresponding reconstructed outputs. The model achieved an average mean squared error (MSE) of 21.5 and an average peak signal-to-noise ratio (PSNR) of 15.4, underscoring its capability to generalize effectively to previously unseen data and validating its practical applicability in real-world scenarios.

\subsection{Experimental results of hybrid optical-digital neural network}
\label{subsec:datasets}
\begin{figure}[htbp]
    \centering
    \includegraphics[width=0.8\textwidth]{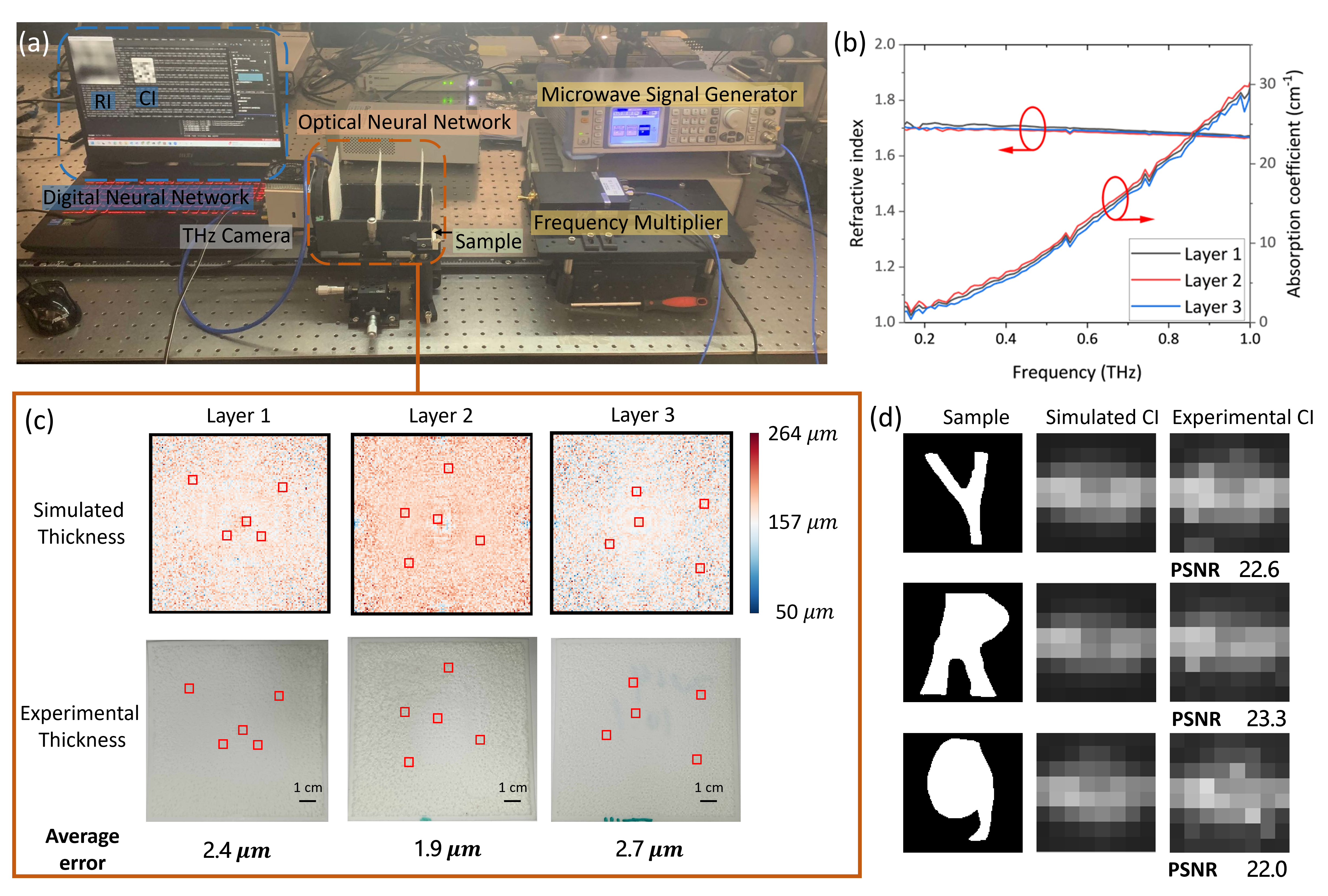}
    \caption{(a) Experimental setup of the hybrid optical-digital neural network THz imaging system. The THz source consists of a microwave signal generator (f = 16.7 GHz) connected to a 12-fold frequency multiplier and a THz horn antenna. The emitted THz wave passes through the sample before reaching ONN, which is composed of three 3D-printed diffractive layers. A THz bolometric sensor array records the compressed THz intensity signal and converts it into a digital format. This digital signal is subsequently processed by a DNN, highlighted in the blue box on the laptop screen, which displays the compressed image (CI) and reconstructed image (RI). (b) The measured absorption loss and refractive index of the three 3D-printed diffractive layers. (c) Comparison of the simulated and the fabricated diffractive layers. (d) Comparison of the ground truth images, simulated compressed images, and experimental compressed images of the “y” “R” and “g” samples.}
    \label{fig:Experimental_setup}
\end{figure}

~\cref{fig:Experimental_setup}(a) represents the experimental setup of the hybrid optical-digital neural network THz imaging system. The THz continuous-wave source comprises a microwave signal generator (model SMB100A, Rohde \& Schwarz), coupled with a frequency multiplier (model WR5.1-VNAX, Virginia Diodes) and equipped with a conical horn antenna (model WR-5.1, Virginia Diodes) to deliver a highly directional THz beam with a 10 dBm emitting power.
This arrangement produces coherent radiation at 0.2 THz by setting the microwave signal generator to 16.7 GHz and subsequently multiplying the frequency twelvefold via the frequency multiplier.

Our ONN is specifically designed to interact with far-field THz waves. Positioned 15~cm away from the THz source---equivalent to approximately \(100 \times 1.5~\mathrm{mm}\) (100 $\lambda$)---this distance satisfies the far-field condition, ensuring accurate beam propagation analysis using the ASM.
The trainable weights embedded within the three diffractive layers of the ONN are optimized through the training process, as described in Section~\ref{subsec:training process}. These layers were fabricated using a masked stereolithography (MSLA) 3D printer (Phrozen, Mighty 4K) with a 20/52-µm axial/lateral resolution. This high-resolution printing allows precise implementation of the trained weight matrices across the diffractive layers.

The fabrication material consists of a low-absorption-coefficient, high-refractive-index composite resin suitable for THz frequencies. This material is formulated from methacrylate oligomers embedded with TiO$_{2}$ nanoparticles~\cite{latifii2024high}. Additionally, an offset thickness was incorporated into all diffractive layers to improve mechanical stability without compromising phase-based operations. To validate the material properties of the 3D-printed THz composite resin, we employed a THz time-domain spectroscopy (THz-TDS) system to measure its absorption and dielectric characteristics. As shown in~\cref{fig:Experimental_setup}(b), the material exhibits a mean refractive index of 1.70 and a mean absorption loss of 3.5~$\mathrm{cm}^{-1}$ across the three diffractive layers. 

A comparison between the simulated and fabricated diffractive layers is presented in~\cref{fig:Experimental_setup}(c). Randomly selected regions were analyzed for printing accuracy, revealing an average dimensional discrepancy of 2.33~$\mu\mathrm{m}$.
To assess the ONN's capability to manipulate THz beam characteristics, three sample patterns (y, R, g) were fabricated and tested in an experimental setup. Their compressed imaging results were first calculated in a simulated environment and then validated using the fabricated ONN in a real-world setting. The results, shown in~\cref{fig:Experimental_setup}(d), demonstrate a strong correlation between experimental and simulated compressed images, achieving an average PSNR of 22.63~dB.

\begin{figure}[htbp]
    \centering
    \includegraphics[width=0.7\textwidth]{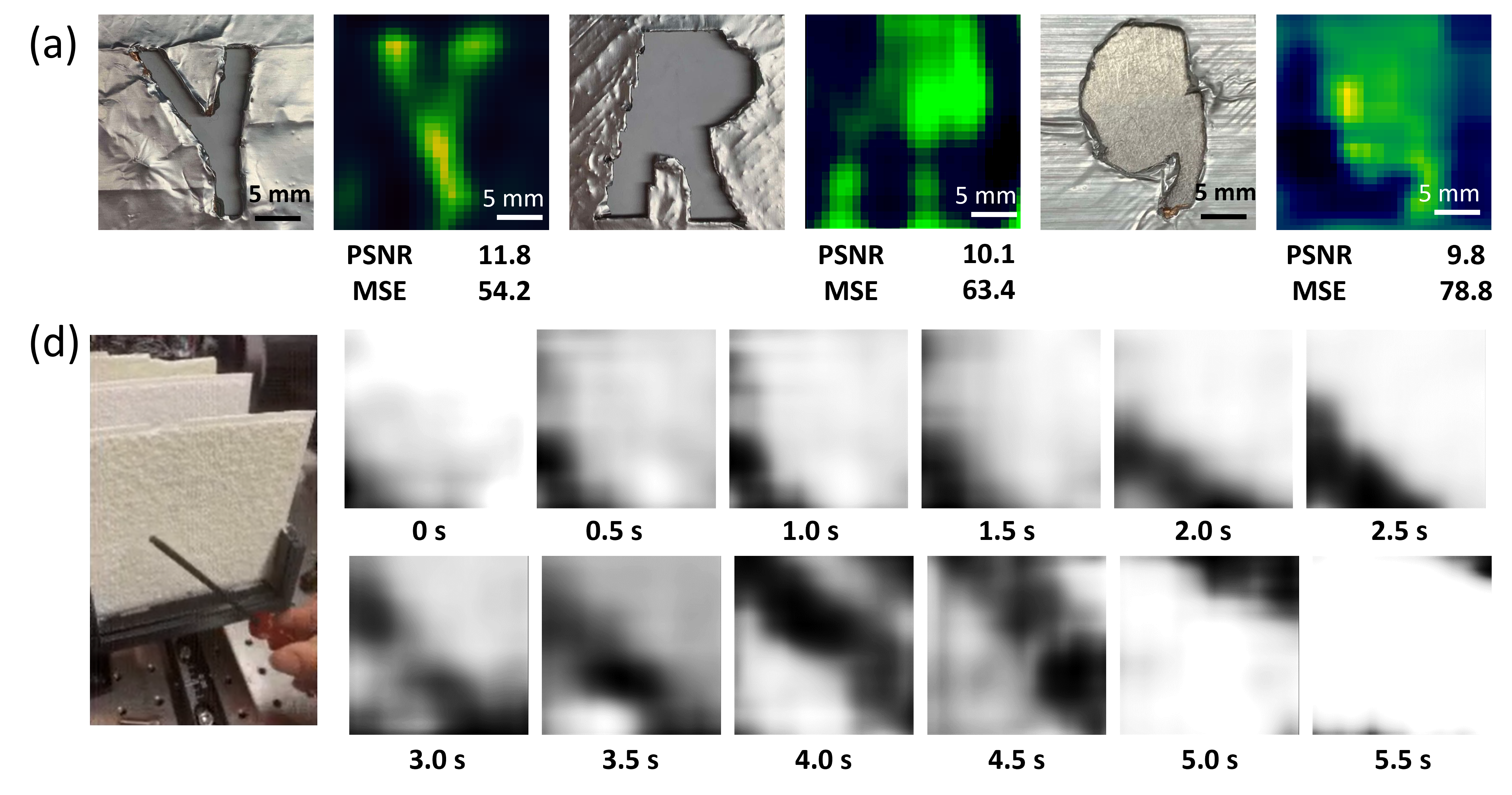}
    \caption{(a) Reconstructed object information obtained from a hybrid optical-digital neural network THz imaging system. The left column showcases optical images of the sample objects, while the right column displays the corresponding reconstructed THz images. (b) Real-time reconstructed THz images of a moving screwdriver, captured with a temporal resolution of 0.5 seconds.}
    \label{fig:Experimental_Result}
\end{figure}

After the THz beam passes through three diffractive layers, the amplitude and phase information is compressed into intensity form and captured by a THz bolometric sensor array with a resolution of \(384 \times 288\) pixels and a total area of \(1.344 \times 1.008~\text{cm}^2\). Despite being the most mature and commercially available option, the THz bolometric sensor array suffers from time-varying thermal drift issues ~\cite{dai2011thermal, li2019thermal, santo2004residual}. To mitigate this, we applied digital processing methods, including average pooling and downsizing of the captured THz bolometric images to \(8 \times 8\) pixel, to effectively reduce the impact of thermal drift.
That means the ONN compresses the dimensionality of the image from $1024 \times 2$ (pixel $\times$ dimension, phase, and amplitude) to  $64 \times 1$ (pixel $\times$ dimension, intensity). Subsequently, the THz sensor array converts these signals into digital intensity form. Normalization is then applied to ensure consistency and compatibility with subsequent processing steps.
The compressed intensity data is then fed into a DNN, whose framework follows the structure outlined in~\cref{subsec:simulation experiments}.
Serving as the decoder, the DNN is jointly trained with the ONN to ensure seamless integration.
The jointly trained DNN interprets the encoded signals from the ONN, decoding them to perform high-dimensional data transformations.
The DNN processes and decodes the compressed imaging data to produce an output at \(128 \times 128\) pixels.
In~\cref{fig:Experimental_Result}, we present the experimental results of the hybrid ONN-DNN THz imaging system.
To validate the system, we used three distinct letters: ``y,'' ``R,'' and ``g.'' The reconstructed images depicted in~\cref{fig:Experimental_Result}(a) exhibit high fidelity and are free of noticeable diffraction artifacts, effectively demonstrating image resolution at the sub-wavelength scale.
The average PSNR and MSE of the experimentally reconstructed images are 10.5 dB and 65.5, respectively.
Additionally, the system successfully covers a \(2.4 \times 2.4~\text{cm}^2\) field-of-view, which is four times larger than the coverage area of the sensor array, without any lensing system in use.

By leveraging the video-rate capability of the THz bolometric sensor array, our imaging system achieves real-time imaging at a frame rate of 
2 fps. It is important to note that this demonstrated frame rate is primarily constrained by our current hardware setup. Integrating advanced graphics processing unit (GPU) modules could significantly enhance processing speed, enabling higher frame rates in real-time imaging capabilities.
To demonstrate the real-world performance of our system, we conducted an experiment using a 5-mm-thick screwdriver as a dynamic sample. The screwdriver was moved at various angles, as shown in~\cref{fig:Experimental_Result}(b) and Supplementary Video 1(a), to assess the system's real-time imaging response. In Supplementary Video 1(b), the results clearly display a distinct line that shifts in correspondence with the screwdriver's movement, illustrating the system's responsiveness and accuracy.
To validate that this observed behavior is a direct result of the hybrid ONN-DNN architecture, we repeated the experiment without the ONN module. In Supplementary Video 1(c), the results reveal an inability to decode any coherent structure or discernible object, underscoring the critical role of the ONN in enabling effective THz image encoding and decoding. These experiments demonstrate not only the real-time imaging capability of the proposed hybrid ONN-DNN system but also its robustness and reliability in capturing and interpreting dynamic THz imaging scenarios.

\section{Conclusions}
\label{sec:Conclusion}
We proposed a hybrid ONN-DNN framework designed to transform signal dimensions, enabling high-speed data acquisition and real-time information reconstruction for THz imaging. In this framework, the ONN serves as an optical encoder, compressing full-dimensional THz signals into a lower-dimensional representation as optical signal carriers propagate through the ONN. Specifically, the ONN transforms the signal from the amplitude-phase domain into the intensity domain. Spatially, the pixel count is reduced from 1024 to 64, allowing high-frame-rate THz FPA sensors to efficiently capture compressed, low-dimensional signals in digital format. The DNN acts as a digital decoder, decompressing and reconstructing the compressed and potentially distorted THz signal back to its original dimensions. This is achieved through a layered convolutional architecture designed to extract object information faithfully. A key innovation of this computational framework is that the ONN not only performs the bulk of computational processing at light speed but also encodes the full object information into an easily recordable latent space. This approach significantly enhances both data acquisition and image reconstruction speeds, meeting the stringent requirements of real-world THz imaging tasks. 

To effectively train both the ONN and DNN, we developed a digital environment capable of simulating the physical behavior of THz beam propagation and light-matter interactions between the THz beam and ONN diffractive layers. Variational autoencoder training methodology was employed to jointly optimize the encoder and decoder functionalities 
For efficient real-time imaging, the ONN was constructed using three diffractive layers, while the DNN comprises 10 convolutional layers, each with 32 channels and an activation function. Transitioning the ONN from a digital to a physical implementation required developing a custom 3D-printable resin composite material with excellent dielectric properties, including a refractive index of 1.70 and an absorption coefficient of 3.5~$\mathrm{cm}^{-1}$ at 0.2~THz. Using a high-precision MSLA 3D printer, we fabricated the physical ONN with deep sub-wavelength-scale lateral and axial resolutions. This approach ensured near-identical geometry and functionality between the digital design and the fabricated ONN. 
The compressive ONN-DNN framework successfully demonstrated its capability to transform signal dimensions and reconstruct high-quality THz images. Clear letter images were reconstructed with an average PSNR of 10.5~dB and an MSE of 65.5. The synergy between the ONN's high-speed signal processing, the THz camera's rapid data recording capabilities, and the DNN's efficient data reconstruction allowed us to achieve real-time imaging at 2 frames per second on a GPU-free laptop.

While this study primarily focuses on improving imaging quality, the hybrid optical-digital compression method also facilitates transforming physical signals into lower-dimensional representations. This transformation empowers the jointly trained DNN to extract valuable information about object interiors, including material properties, geometry, chemical composition, physical characteristics, and molecular dynamics. Furthermore, the ONN-DNN framework is flexible and can integrate alternative architectures, such as residual networks~\cite{shafiq2022deep} or transformer-based models~\cite{khan2022transformers}. This adaptability allows for enhanced efficiency, effectiveness, and customization to meet specific application objectives. Importantly, the proposed compressive ONN-DNN platform is not limited to the THz band. It has the potential to operate across other regions of the electromagnetic and acoustic spectra, including radio waves (RF), millimeter waves (mmWave), infrared (IR), visible light, and ultrasonic bands. These domains share fundamental principles of wave propagation and wave-matter interaction, making the framework applicable to communication, sensing, and imaging tasks. In essence, this hybrid approach challenges conventional imaging methodologies and opens new avenues for rethinking system architecture, data utilization, signal processing, and information extraction in optical and acoustic systems.



\begin{backmatter}
\bmsection{Funding}National Science and Technology Council, Taiwan (113-2628-E-007-011-MY3).

\bmsection{Acknowledgments}
The authors thank Mr Pierre Talbot (INO) and their NTHU colleagues, including Professors Yi-Chun Liu and Kai-Ming Feng , along with fellows Ms Yun-Ting Tseng, Mr Tsung-Han Wu, Mr Yi-Chun Hung, Mr Pouya Torkaman, Mr Han-Yu Liang, Mr Chun-Yu Kao, and Mr Chia-Ming Mai for their valuable support and insightful suggestions. 

\bmsection{Disclosures}The authors declare no conflicts of interest. 

\bmsection{Data Availability}Data underlying the results presented in this paper are not publicly available at this time but may be obtained from the authors upon reasonable request. 

\end{backmatter}

\bibliography{2_refs.bib}

\end{document}